# From metagenomic data to personalized computational microbiotas: Predicting dietary supplements for Crohn's disease

Eugen Bauer[1] and Ines Thiele[1,2,*]

[1]Luxembourg Centre for Systems Biomedicine, Universite du Luxembourg, Esch-sur-Alzette, L-4362, Luxembourg

[2]Lead Contact

[*]Correspondence: ines.thiele@uni.lu

## Summary

Crohn's disease (CD) is associated with an ecological imbalance of the intestinal microbiota, consisting of hundreds of species. The underlying complexity as well as individual differences between patients contributes to the difficulty to define a standardized treatment. Computational modeling can systematically investigate metabolic interactions between gut microbes to unravel novel mechanistic insights. In this study, we integrated metagenomic data of CD patients and healthy controls with genome-scale metabolic models into personalized *in silico* microbiotas. We predicted short chain fatty acid (SFCA) levels for patients and controls, which were overall congruent with experimental findings. As an emergent property, low concentrations of SCFA were predicted for CD patients and the SCFA signatures were unique to each patient. Consequently, we suggest personalized dietary treatments that could improve each patient's SCFA levels. The underlying modeling approach could aid clinical practice to find novel dietary treatment and guide recovery by rationally proposing food aliments.

# Introduction

The human gut microbiota is composed of thousand different bacterial species with a large functional diversity that surpasses the human genome in terms of the collective gene pool (Qin et al., 2010). Health promoting functions of the gut microbiota include the breakdown of otherwise indigestible dietary fibers and production of short chain fatty acids (SCFA) utilized by the human host (den Besten et al., 2013b).

Various human diseases, including inflammatory bowel disease (IBD), are associated with a loss of functional and taxonomic diversity of the gut microbiota (Qin et al., 2010). The main symptom of IBD is inflammation of the gut epithelium (Khor et al., 2011). Depending on the site of inflammation, IBD is distinguished in ulcerative colitis, which primarily affects the colon, and Crohn's disease (CD), in which different sites can be affected. Non-invasive treatments for CD include the intake of antibiotics (Prantera et al., 1996) and steroid therapies, which suppress the immune system (Van Dullemen et al., 1995). In addition, dietary change with a defined formula is used to ease the symptoms of the disease (Wilschanski et al., 1996). However, the success of these diet formulas varies between patients (Griffiths et al., 1995). Additionally, after remission, patients have difficulties in finding an appropriate diet and often experience relapse. Considering the metabolic relevance of the human gut microbiota, it has been suggested that the dietary formula effects and reshapes the microbiota (Kaakoush et al., 2015). Overall, the microbial diversity is decreased in CD patients. A shortage of SCFAs (Huda-Faujan et al., 2010) coincides with a decreased abundance of fermenting Firmicutes bacteria (Manichanh et al., 2006). Microbial SCFAs have been recognized as important modulators of the immune system and as a nutrition source (Guarner and Malagelada, 2003). Butyrate, for example, is taken up as an additional energy source by the host (Donohoe et al., 2011), contributes to epithelial barrier integrity (Peng et al., 2007), and stimulates the immune system (Furusawa et al., 2013). CD patients suffer from a low butyrate concentration (De Preter et al., 2013), but its dietary supplementation can revert many of the IBD symptoms (Sabatino et al., 2005), highlighting the relevance of this particular SCFA in CD.

Given that the human gut microbiota is a complex microbial community with many different microbes that have varying metabolic potentials and substrate affinities (Bauer et al., 2015), it becomes difficult to track the manifold ecological interactions that differ between CD patients and healthy individuals. Meta-omics approaches are generally used to characterize the microbiota and its metabolic potential (Zoetendal et al., 2008). However, these top-down approaches do not provide mechanistic insights on the resilience of the microbiota and how perturbations, such as dietary treatments, may affect the system as a whole.

Bottom-up systems biology approaches can mechanistically describe biological systems and make biologically relevant predictions. In particular, constraint-based reconstruction and analysis (COBRA) has been successfully applied to model the metabolism of different species and make predictions on how perturbations affect the metabolic phenotype (Orth et al., 2010). Briefly, genome-scale metabolic reconstructions are represented by the complete set of biochemical reactions derived from a genome annotation and organism-specific literature in a stoichiometric accurate manner (Thiele and Palsson, 2010). Such high-quality manually-curated metabolic reconstructions are available for organisms from all three domains of life, such as *E. coli* (O'brien et al., 2013), yeast (Nookaew et al., 2011), and human (e.g., (Thiele et al., 2013b)). Through the application of specific constraints (e.g., nutrient availability), the metabolic reconstructions can be converted into condition-specific models, which predict the reaction flux rates and growth yield under a given objective that is optimized using flux balance analysis (FBA) (Orth et al., 2010). In a recent publication (Bauer et al., 2017), we combined FBA with agent based modeling to simulate the ecology of microbial communities through the BacArena framework. Based on dietary metabolites in the environment, the metabolic models are constrained to simulate the metabolic states of each species, optimizing their growth yield. Metabolic interactions emerge from the exchange of metabolites between species and the environment. These interactions can influence the metabolite concentration and the microbial community by inducing cross-feeding or resource competition. Such COBRA-based approaches provide a powerful mean to investigate mechanistic links in complex biological systems, such as the human gut microbiota.

A recent study on pediatric CD sequenced the metagenomes of a North American cohort consisting of 26 healthy controls and 85 patients newly diagnosed with CD (Lewis et al., 2015). In their study, the authors could distinguish two clusters of patients: A cluster of 57 patients, which had a microbiota composition similar to the healthy controls, and a cluster of 28 patients that had a distinguished dysbiotic microbiota. Compared to controls, these dysbiotic patients had a strongly differing functional and microbial abundance profile.

Here, we retrieved the original metagenomic data of the 26 healthy controls and 28 dysbiotic patients (Lewis et al., 2015) to simulate personalized *in silico* microbiotas with BacArena. We demonstrate that the simulated metabolic differences between patients and controls are congruent with experimental findings. We further show that predicted individual specific SCFA signatures are unique to each patient. Based on these results, we then predict personalized dietary treatments that would improve the SCFA concentrations of each patient. With this work, we demonstrate the added value of performing computational modeling in conjunction with high-throughput data of individual microbiotas to predict mechanism-based personalized dietary intervention strategies for CD patients.

# Results

The aim of the present study was to predict *in silico* novel personalized dietary treatments for CD and investigate individual differences. We simulated personalized *in silico* microbiotas consisting of hundreds microbial metabolic models as defined by published metagenomic data of healthy controls and CD patients (Lewis et al., 2015) using a novel computational modeling approach (Bauer et al., 2017) in which we combined FBA with agent based modeling to simulate the ecology of microbial communities through the BacArena framework. Based on dietary metabolites in the environment, the metabolic models are constrained to simulate the metabolic states of each species, optimizing their growth yield. Metabolic interactions emerge from the exchange of metabolites between species and the environment. These interactions can be used to gain further insight into metabolic differences that may contribute to CD and to propose modeling-assisted dietary intervention strategies for CD patients (Figure 1). We describe differences between healthy controls and CD patients based on SCFAs as well as microbial abundances, which we validated with existing experimental knowledge. Individual differences within patients and controls were assessed to find the SCFA signature specific to each individual microbiota. Based on the individual microbiotas, personalized dietary treatments, such as supplementation of pectin and different glycans, were predicted to equilibrate the SCFA concentrations and promote healthier SCFA concentrations. Taken together, our work demonstrates the use of computational modeling to integrate existing high-throughput data of individual microbiotas and mechanistically predict novel personalized dietary treatments for CD.

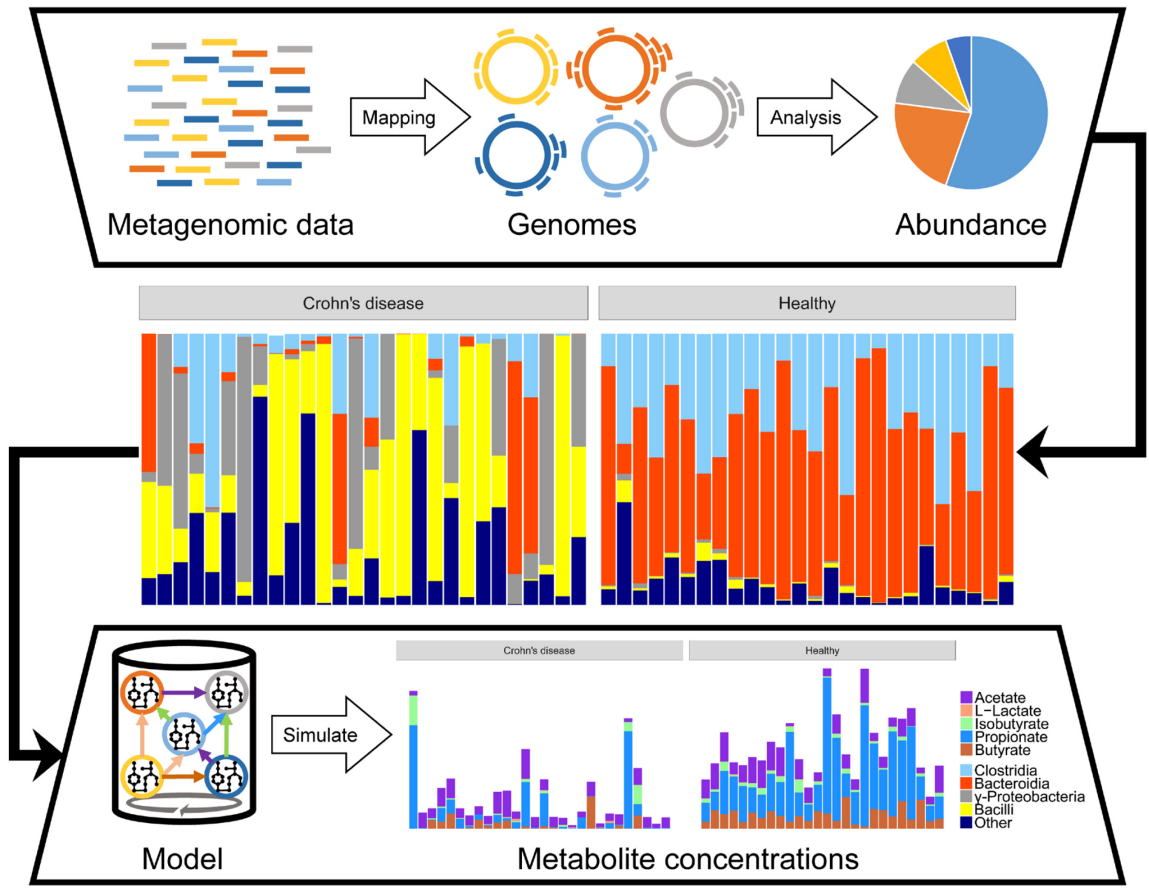

**Figure 1: Computational framework used to create personalized metabolic models of gut microbial communities**. Published metagenomic data were integrated into an *in silico* microbiota model for each CD patient and healthy control to simulate emergent metabolite concentrations.

## Microbial differences between healthy controls and CD patients

We ensured that our computational workflow (Figure 1) would not alter the reported microbial differences between healthy controls and of dysbiotic CD patients (Lewis et al., 2015). The workflow mapped the published metagenomic data of healthy controls and CD patients onto the genome sequences of the 773 gut microbial strains, for which metabolic reconstructions were available (Magnusdottir et al., 2016). In average, 283 +/- 240 of the 773 microbial strains were covered in the personalized *in silico* microbiota (Figure S1). Notably, the smallest *in silico* microbiota contained only eight microbes, while the biggest had 713 of the 773 microbial strains. There were seven out of 54 *in silico* microbiotas that had less than 40 of the 773 microbes. While CD patients had generally less microbes, there were also some healthy controls with less than 40 microbes and some CD patients with more than 600 microbes (Figure S1). Overall, the personalized *in silico* microbiota captured 73.5 +/- 16 % of the relative microbial abundance from the original metagenomic reads. We could observe a clear separation of the healthy controls and CD patients when assessing the microbial differences based on microbial abundances captured by the *in silico* microbiota (Figure 2A), which was independent on the used similarity metrics (Figure S2). The most pronounced differences between the healthy and the CD individuals were due to significantly higher abundance of Bacilli and Gammaproteobacteria (p<0.05, Wilcoxon rank-sum test) and significantly lower abundance of Bacteroidia and Clostridia (p<0.001, Wilcoxon rank-sum test) in CD patients (Figure 2D).

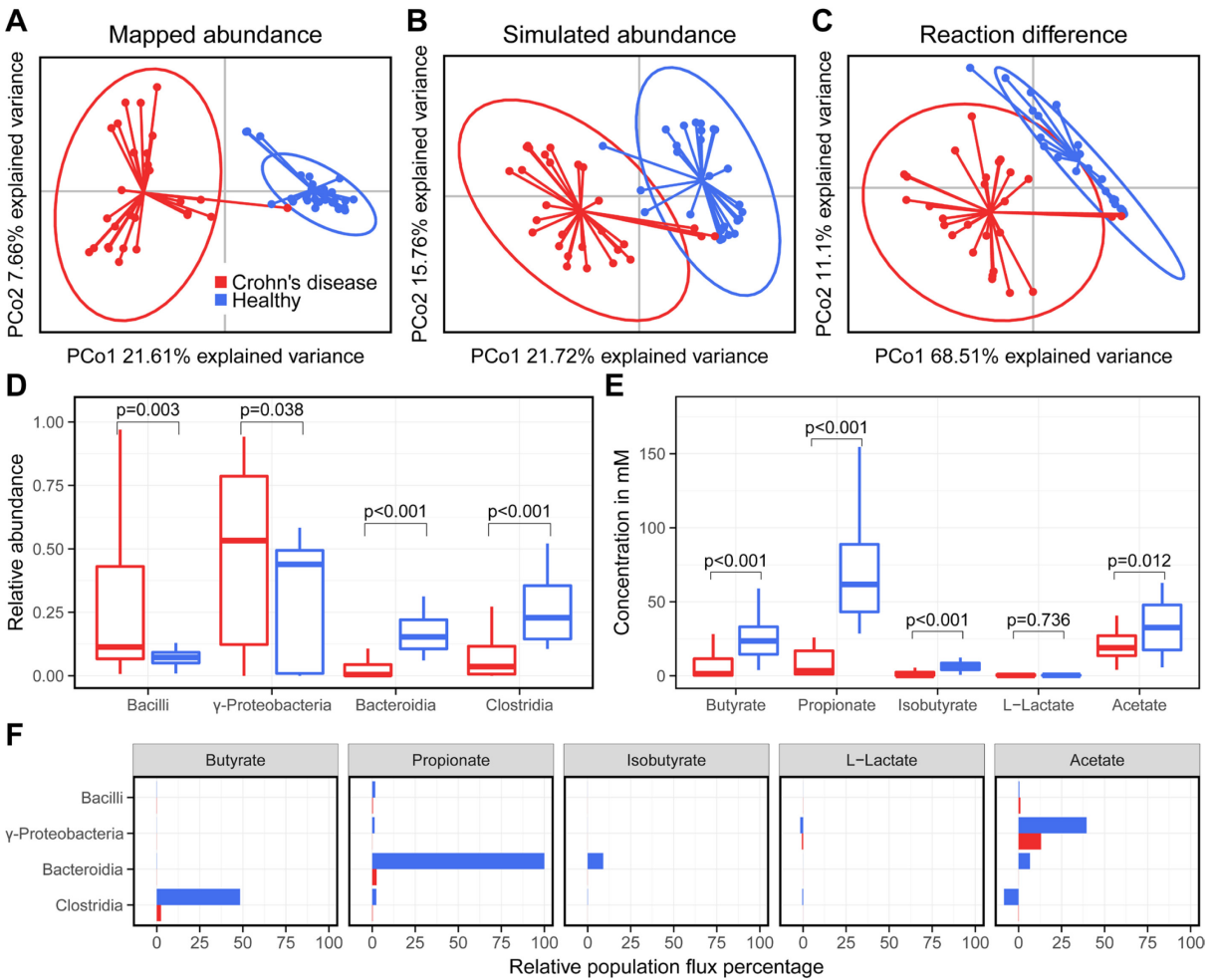

**Figure 2: Metabolic and microbial group variability between healthy controls and Crohn's disease patients**. Similarities were assessed based on a principle coordinate analysis (PCoA) of the reaction content with jaccard distance (A), mapped reads with bray curtis dissimilarity (B), and simulated abundances with bray curtis dissimilarity (C). Based on the simulation, relative abundances (D) and metabolite concentrations of fermentation products (E) were compared (p-value determined by Wilcoxon rank-sum test). Microbial metabolic activities were displayed as the total population flux (F).

We then simulated the personalized *in silico* microbiota, which were inoculated with 500 microbes on a grid with 10,000 cells for 24 hours in the BacArena framework and analyzed whether the microbial abundances changed compared to the initial (metagenomic data driven) abundances. At the end of the simulation, the grid was populated by an average of 5902 +/- 1743 microbes corresponding to an average grid occupation of 59 +/- 17 %. Overall, the simulated abundances recapitulate the initial microbial differences, demonstrating that the *in silico* microbiotas were stable over time in BacArena (Figure 2B). However, the abundance ratios of four out of 28 genera were consistently higher in CD patients based on the simulated abundance, but lower based on the mapped data (Figure 3A). In contrast, the mapped abundance data showed good agreement with the abundances reported in the original study (Figure 3A, Figure S3). This discrepancy can be explained by the CD patients having a lower diversity of microbes, which led to a higher predicted abundance for the present genera.

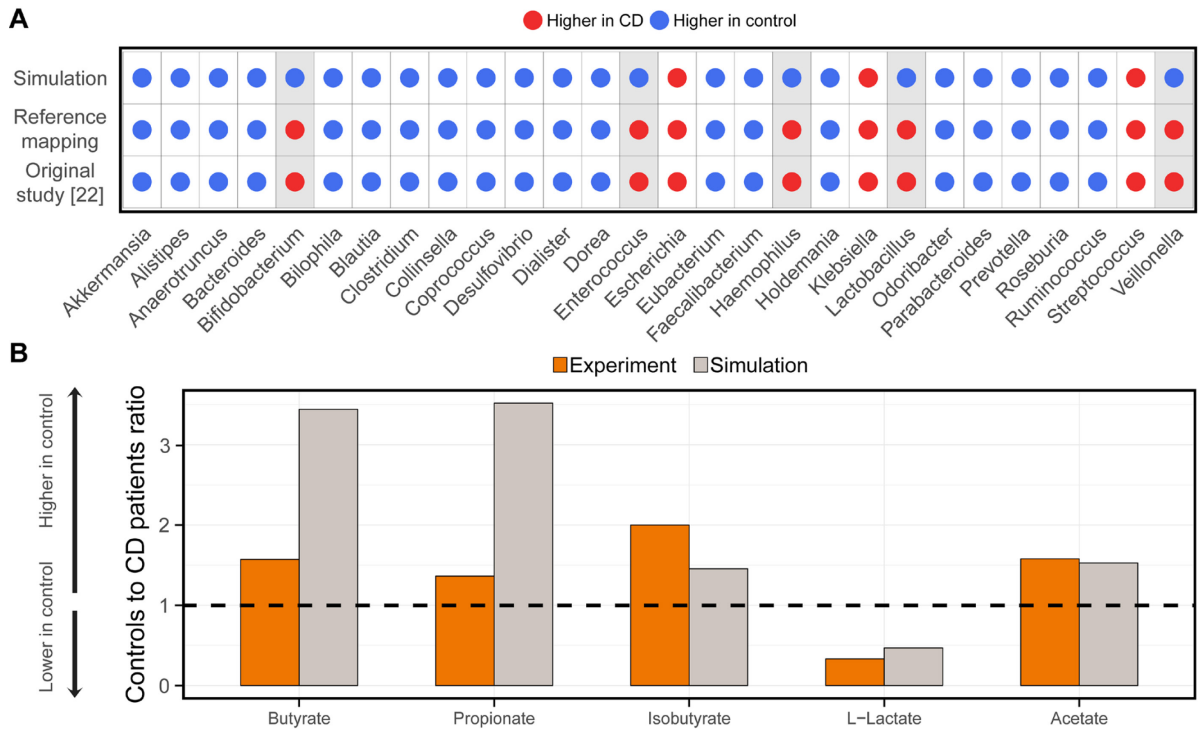

**Figure 3: Qualitative comparison of simulation results with experimental values**. Experimental relative abundances of microbial genera (A) were retrieved from the original study (Lewis et al., 2015) and compared with the abundances based on the mapped reads and simulations (t=24h). (B) Metabolite concentrations were retrieved from an independent experimental study (Hove and Mortensen, 1995) and compared with the simulations (t=24h) based on the mean concentration ratios of healthy controls and CD patients.

Taken together, our workflow recapitulates the reported microbial differences between controls and CD patients (Lewis et al., 2015). Furthermore, the simulation results of the personalized *in silico* microbiota in BacArena illustrate that these microbes can also co-exist, in a similar relative abundance, as stable microbial communities *in silico*.

## Emergent metabolic differences between healthy controls and CD patients

We investigated whether the difference in microbial abundance in the personalized *in silico* microbiota also corresponded to differences in metabolic repertoires. That is: Are there microbes in some of the personalized *in silico* microbiota that are unique and may help to distinguish healthy and CD microbiota? In average, each personalized *in silico* microbiota consisted of 3,332,957 +/- 285,848 metabolic reactions belonging to 3,036 +/- 424 unique metabolic reactions. The presence and absence pattern of the unique reactions in the *in silico* microbiota varied between individuals as well as between the two groups (Figure 2C). Interestingly, based on the reaction content, the first two principal components explained almost 80% of the variation in the data (Figure 2C), and were mainly driven by the presence of transport reactions for fibers (Table S1). The observed reaction based separation is consistent with the aforementioned differences in microbial classes (Figure 2D) and the distinct fiber metabolizing properties of Bacteroides.

SCFAs are important energy precursors and interact with the human immune system (Furusawa et al., 2013). We analyzed the secretion of SCFAs after 24 hours by each personalized *in silico* microbiota, consisting of a myriad microbial models that can individually consume and secrete the SCFAs, to establish whether known microbiota-level differences in SCFA production could be reproduced by our computational modeling approach. The SCFAs butyrate, propionate, isobutyrate, and acetate were significantly lower in CD patients ($p<0.05$, Wilcoxon rank-sum test, Figure 2E). Only L-lactate levels were slightly higher in CD patients. To check for the validity of the simulated metabolite concentrations, we compared our results with an independent experimental study (Hove and Mortensen, 1995). The qualitative difference between CD patients and healthy controls were consistent with our simulations (Figure 3B). However, the predicted concentrations of butyrate and propionate were three times higher in controls than in CD patients (Figure 3B), which is much higher than the reported difference, probably due to the absence of the host cells in our model setup that can take up butyrate and propionate produced by the microbiota (den Besten et al., 2013a). Overall, our results confirm that the personalized *in silico* microbiotas also recapitulate known differences in SCFA production levels in healthy and CD individuals.

An advantage of using computational modeling is that we can determine which microbes, or microbial classes, in the *in silico* microbiota caused the predicted differences in SCFA production. Therefore, we analyzed the summed uptake and secretion fluxes of each microbial class in the personalized *in silico* microbiota. We found that Clostridia were responsible for the production of 50% of the total butyrate, Bacteroidia produced almost 100% of the total propionate and about 10% of the total isobutyrate, Bacilli produced small quantities (<5% of the total concentration) of L-lactate, and Gammaproteobacteria produced almost 50% of the total acetate (Figure 2F). Notably, in healthy controls, acetate was taken up by Clostridia illustrating a cross-feeding mechanism between Gammaproteobacteria and Clostridia. These results demonstrated how changes in representatives of the main microbial classes can result in differences in SCFA production capabilities that differ significantly between healthy controls and CD patients.

## SCFA production profiles are patient-specific

The original metagenomic study (Lewis et al., 2015) reported the most distinct microbial differences between the healthy controls and the CD patients but also variability within the groups. Accordingly, the simulated relative microbial abundance also varied between the individuals (Figure 4, left). We next investigated how much the predicted SFCA production capability varied between CD patients. Two (CD10, CD11) out of 28 CD patients had butyrate levels that were comparable to the mean of healthy controls (mean concentration of 7.5 and 25.8 mM for CD and controls respectively). This could be explained by the higher metabolic activity of Clostridia species in these patients (Figure 4, right). In three cases (CD2, CD4, CD22), the concentration of isobutyrate was higher in CD patients (Figure 4) compared to the mean concentration in healthy controls (mean concentration of 4.9 and 7.1 mM for CD and controls respectively). Two of these patients (CD2, CD22) also had propionate levels comparable to the mean of healthy controls (mean concentration of 25 and 87.9 mM for CD and controls respectively), which is congruent with the high metabolic activity of isobutyrate and propionate producing Bacteroides species (Figure 4, right). Twelve out of the 28 patients showed increased L-lactate concentrations (mean concentration of 0.7 and 0.3 mM for CD and controls respectively), which can be attributed to the metabolic activity of Bacilli and other taxa (Figure 4). This is also congruent with the observation, that CD patients had an overall higher L-lactate concentration (Figure 2E). Five patients (CD11, CD16, CD17, CD19, and CD25) showed acetate levels that were comparable to the mean of healthy controls (mean concentration of 21.1 and 32.2 mM for CD and controls respectively). This can be mostly attributed to the metabolic activity of Bacilli and Gammaproteobacteria (Figure 4, right). Overall, these results indicated that every patient has a specific signature of SCFAs, in which some metabolites have levels comparable to healthy controls. This observation can be explained by the metabolic activity of the present microbiota, indicating that metabolic stimulation of the native CD microbiota may be able to revert some of the differences between CD and healthy controls.

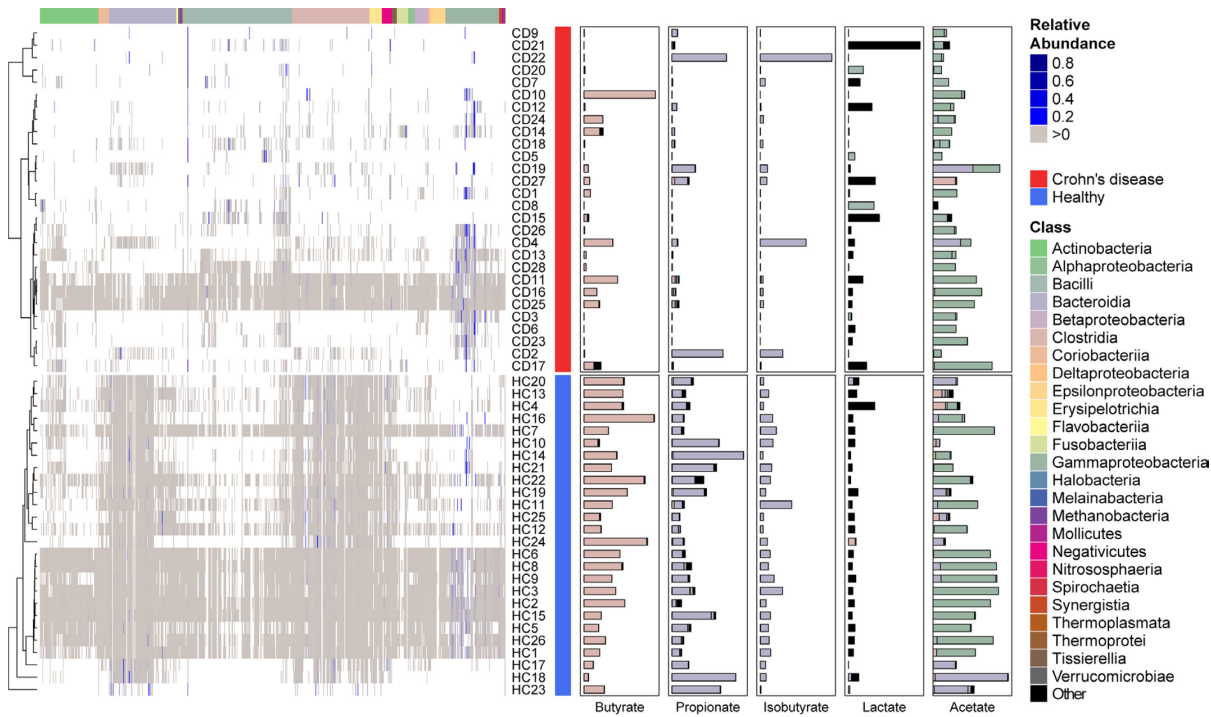

**Figure 4: Individual variability between CD patients and healthy controls.** The presence of different microbes is indicated by a gray color and the relative abundance by a blue color scale. Microbial taxa are based on the class level. Predicted metabolite concentrations are based on simulations. The microbial contribution to the concentrations are based on metabolic fluxes.

# Personalized dietary intervention strategies to normalize SCFA production capabilities of the personalized in silico microbiota

Defined dietary regimes are one possible treatment strategy for CD patients (Wilschanski et al., 1996). However, the success of this treatment varies between CD patients (den Besten et al., 2013a). As butyrate is often low in CD patients and its dietary supplementation has been reported to ease gastrointestinal symptoms of CD patients (Sabatino et al., 2005), we investigated whether we could design personalized dietary interventions that would restore the SCFA production to levels commonly reported in healthy individuals. We approached this problem by predicting first whether increasing each dietary compound, present in the *in silico* rich diet, could individually lead to a more healthy level of each of the five SCFAs in any microbial model present in a given CD patient *in silico* microbiota (Figure 5A). Interestingly, the number of the predicted dietary metabolites to be supplemented was highly specific for each patient and ranged between 1 and 55 metabolites (median of 19 metabolites) (Figure 5B). For four out of the 28 CD patients, our described prediction approach did not identify any metabolites that could normalize the production of any SCFA. These four patients had a higher abundance of Gammaproteobacteria and Bacilli, while major SCFA producers, belonging mostly to Bacteroidetes and Clostridia were largely absent. For the remaining 24 CD patients, the most prominent category of the predicted metabolites were mucus glycans and glycosaminoglycans (Figure 5B). In particular, pectin supplementation was predicted to be a good dietary supplement for 17 out of 24 CD patients (Figure S4). Other prevalent metabolites included various specific human produced mucus glycans and hepan/hyaluronan proteoglycan degradation products as well as plant-derived larch arabinogalactan, lavanbiose, and amylose.

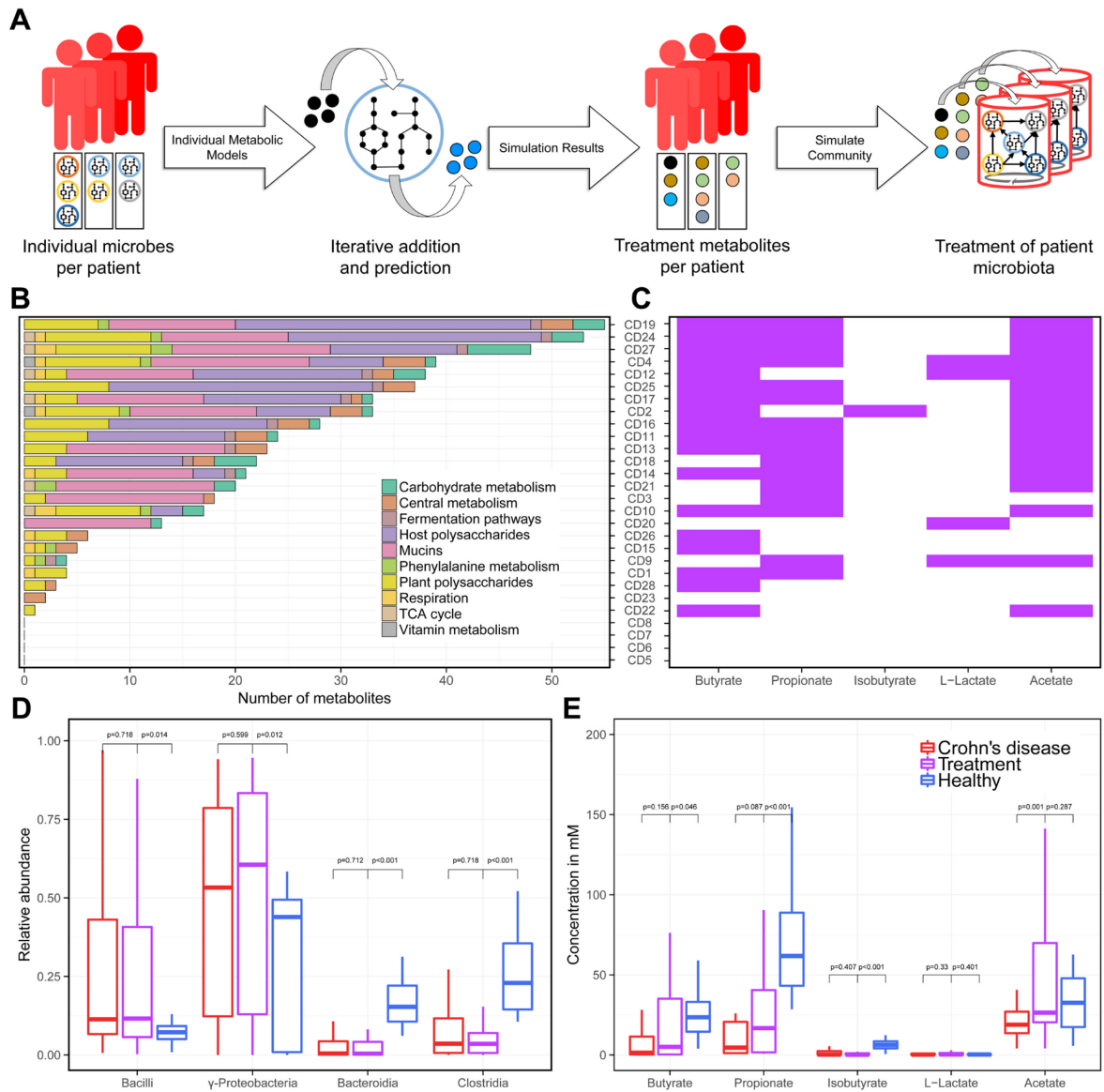

**Figure 5: Individual treatment prediction for each CD patient.** For the prediction of treatment metabolites (A), single metabolic models of microbes for each patient were optimized for the production of the target metabolites with iterative dietary additions. Panel (B) shows broader categories of the predicted metabolites and (C) shows the response (metabolite increase of 25%) of each patient in purple. Panel (D) and (E) show the relative abundance and metabolite concentrations.

We then added all of these identified metabolites to each of the personalized *in silico* microbiota to ensure that the community could also produce healthier SCFA levels. Each personalized *in silico* microbiota was simulated for 24 hours in the personalized supplemented diets. The success of the *in silico* dietary interventions varied between the patients (Figure 5C). Overall, the most successful individual metabolite level restoration was obtained for butyrate, propionate, and acetate, whereas the *in silico* treatment was less successful for isobutyrate and L-lactate (Figure 5C). Overall, the SCFA concentration differences between CD patients and healthy controls (Figure 5D) were also improved with respect to butyrate, propionate, and acetate, whereas isobutyrate and L-lactate did not show improved levels compared to the untreated concentrations (Figure 5D). The *in silico* treatments had only small effects on the relative species abundances in the personalized *in silico* microbiota as they were not able to restore a more healthy ratio of microbial species (Figure 5E) due to the dysbiotic patients lacking the relevant microbes found in healthy individuals. Therefore, our results showed quantitatively improved levels of SCFAs on the individual patient level as well as on the differences between patients and healthy controls. To re-establish a healthier microbiota composition, we would need to account for food-borne microbes or probiotics in addition to the dietary treatment.

## Discussion

We created personalized *in silico* microbiota of healthy controls and CD patients by integrating metagenomic data into a bottom-up systems biology framework (Figure 1). Recent approaches have successfully integrated metagenomic data to model the ecological dynamics of the human gut microbiota (Bashan et al., 2016) but lack the metabolic aspect, which plays an important role for human health and disease (Tremaroli and Bäckhed, 2012). Therefore, the added benefit of our modeling approach is that we combine metabolism and ecology to investigate the metabolic activity of the human gut microbiota.

To find strong differences between CD patients and healthy controls, we selected data of dysbiotic patients that were defined by their microbial distance to healthy controls (Lewis et al., 2015). Expectedly, we could reproduce the microbial differences originally reported in the study (Figure 2A). Moreover, our reference based assessment was consistent with the reference independent analysis in the original study (Figure 3A), which further demonstrates that the set of 773 AGORA microbes capture the most common human gut microbes (Magnusdottir et al., 2016). When comparing the abundance of specific genera (Figure 3A), the community simulations predict differing ratios for four out of 28 genera, which indicates a minor variability in the simulations that did not affect the overall differences (Figure 2B). The main microbial differences between CD patients and healthy controls can be attributed to a decreased abundance of Bacteroidia and Clostridia as well as an increased abundance of Bacilli and Gammaproteobacteria in CD patients (Figure 2D), which was in accordance with an independent experimental study (Kaakoush et al., 2012) and characteristic for a dysbiotic microbiota. Our results therefore capture strongly dysbiotic CD microbiotas, which is, however, only one specific case of CD (Lewis et al., 2015). This approach allows us to address fundamental questions in CD dysbiosis and to mechanistically describe how the microbiota can shape metabolite concentrations, which is less understood so far.

The simulated SCFA concentrations represent emergent properties of our models that could not be achieved by the metagenomic data alone. Therefore, we could simulate clinical relevant metabolite concentrations, known to be differentially regulated in CD (Hove and Mortensen, 1995). Interestingly, we could see differences in SCFA levels between healthy controls and CD patients (Figure 2E). In particular, higher concentrations of acetate, propionate, butyrate, and isobutyrate as well as a lower concentration of L-Lactate in controls (Figure 2E). We could also validate these qualitative differences with experimental literature (Hove and Mortensen, 1995) (Figure 3B). Based on the quantitative ratios between controls and patients, butyrate and propionate were higher in our simulations than in experiments (Figure 3B). This apparent discrepancy could be explained by the uptake of butyrate and propionate by the host (den Besten et al., 2013a), which we did not include and is therefore a limitation of our current modeling approach. SCFAs, in general, have been associated with healthy gut functions, such as energy conversion of the host as well as immune stimulation (Guarner and Malagelada, 2003). Butyrate, in particular, mediates the immune system (Furusawa et al., 2013) and influences the tight junctions between epithelial cells (Peng et al., 2007). Moreover, butyrate, as well as propionate, are carbon sources for colonocytes (Clausen and Mortensen, 1995; Roediger, 1982). Taken together, the added value of our modeling approach is that we can predict these qualitative changes in SCFA levels, which we can attribute to specific microbial metabolic activity.

We identified which microbes are responsible for the production of the SCFA (Figure 2F). Clostridia produced mainly butyrate explaining its lower concentration in CD patients (Figure 2E), who had generally lower Clostridia abundances (Figure 2D). The Clostridia Faecalibacterium and Roseburia are known to be the main butyrate producers (Machiels et al., 2013). Interestingly, these two Clostridia were decreased in abundance in CD patients (Figure 3B). We identified novel metabolic interaction patterns, such as the consumption of acetate by Clostridia (Figure 2F). *In vitro* experiments have demonstrated cross-feeding interactions between Clostridia and Bifidobacterium species (Belenguer et al., 2006). Based on our simulation, we suggest that these interactions take also place in an ecologically complex microbiota and that they are of high relevance in the human gut. These metabolic interactions link microbes with metabolites and demonstrate that we capture *in silico* the gut microbiota as a whole.

Our personalized *in silico* microbiota modeling approach permitted the investigation of individual differences between CD patients and healthy controls (Figure 4). Overall, we found that healthy controls have a higher diversity of microbes than CD patients, which is also confirmed by experimental knowledge (Manichanh et al., 2006). Consequently, controls have more comparable SCFA levels (Figure 4), indicating metabolic consistency through functional redundancy of microbes (Human Microbiome Project Consortium, 2012). Interestingly, some patients had some SCFA but not all levels comparable with the controls, which could be attributed to a higher metabolic activity of health relevant microbes in those patients (Figure 4). One could speculate that the microbiota of CD patients can compensate some metabolic differences but lacked functional redundancy and diversity to consistently establish a healthy SCFA signature (Figure 4). This observation further underlines the importance of a diverse microbiota, which can complement potential metabolic shortcomings between microbes and consistently produce SCFAs. Further studies could investigate the importance of keystone species in this context, which have a low abundance but high metabolic activity and thus ecological relevance (Trosvik and Muinck, 2015).

In our *in silico* treatment predictions, we take the individual factors into account by designing dietary supplements that would compensate the individual differences (Figure 5A). Most of the predicted treatment metabolites were mucus glycans, glycosaminoglycans, and plant polysaccharides (Figure 5B), further indicating that fibers are relevant in shaping the gut microbiota metabolism (Koropatkin et al., 2012; Maxwell et al., 2012). Particularly, pectin was predicted as a potential treatment for the majority of patients, which further points towards the dietary relevance of this compound (Maxwell et al., 2012). Plant fibers and host glycans can influence the gut microbiota by stimulating Clostridia and Bacteroidia species (Flint et al., 2008), which produce butyrate and propionate, respectively (Figure 2F). Interestingly, the predicted metabolite cocktails were different for each patient (Figure 5B, Figure S4). In clinical practice, a standard dietary formula in form of exclusive enteral nutrition is used to treat patients with CD (Wilschanski et al., 1996). However, not every patient responds equally well to different diet formulations, which vary in their fiber content (Lien et al., 1996). Current knowledge is limited when defining personalized diets because of the complexity of the human gut microbiota and its intricate response to different diets. Some patients suffer from relapse when switching to a normal diet after successful remission (Belluzzi et al., 1996). In such cases, our modeling-based predictions could give novel directions on aliments based on a patient's microbiota. Finally, a dietary treatment strategy for an individual is not likely to be static. Using computational modeling in conjunction with metagenomic data, the dietary treatment could be readily redefined and adjusted to match the patient's need. To our knowledge, such modeling-guided dietary treatment approach is not available yet for Crohn's disease patients. As a next step, our predictions need to be validated in a nutritional trial. Then, our systematic approach to defining personalized nutrition therapies could guide clinicians and nutritionists in designing new, personalized diet-based treatments.

Testing our *in silico* dietary treatments on each patient's' microbiota, we found an improvement in SCFA levels. Butyrate, propionate, and acetate showed an overall success in shifting levels, while isobutyrate and L-lactate were less successful (Figure 5C, 5E), since those SCFAs only had a minor difference between controls and patients (Figure 2E). The overall microbe abundance did also not shift significantly in the treatment condition (Figure 5D), because patients had a lower diversity from the start (Figure 4) and could not acquire the necessary microbes to compensate their abundance profile. Further studies could simulate the effect of adding specific microbe models as a treatment, which could be integrated in our framework. Furthermore, human metabolism could be integrated with the *in silico* microbiota to investigate the reciprocal effect on the host, and, for instance, the effect of colorectal cancer cells that might be affected by butyrate concentrations (Sengupta et al., 2006).

Several studies emphasize the need for computational models to discover novel hypothesis and mechanisms for microbiota associated diseases (Biggs et al., 2015; Ji and Nielsen, 2015; Thiele et al., 2013a). Our approach introduces metabolism as an additional emergent property of the microbiota yielding into novel mechanistic insight of SCFA production by microbial communities. So far, our approach ignores for simplicity the spatial component, which could have important implications in shaping the ecology and possible metabolic interactions. An extension for possible treatment strategies includes the simulation of probiotics and fecal transplantation. In fact, our model could be used as an additional workflow for donor optimization of fecal transplantation (Pamer, 2014) by finding the most appropriate microbiota. Most importantly, the computational modeling approach that we presented here is not limited to the application of CD but can be applied to any metagenomic data set and disease with microbial dysbiosis. Taken together, we present a powerful, expandable, versatile computational modeling approach that permits to yield novel insight into metabolic interactions emerging from personalized metagenomic data and to predict personalized dietary intervention strategies.

# Author Contributions

I.T. and E.B. designed and conducted the study. E.B. performed simulations and analyzed data. I.T. and E.B. wrote and edited the manuscript.

# Acknowledgements

We want to thank Dr. Almut Heinken for classifying the treatment metabolites and giving useful comments on the analysis of the results. We also want to thank Mr. Marouen Ben Guebilla, Mr. Alberto Noronha, and Mr. Federico Baldini for giving useful comments on the manuscript. This work was supported by an ATTRACT program grant (FNR/A12/01), and an Aides a la Formation-Recherche (FNR/6783162) grant.

# References

Bashan, A., Gibson, T.E., Friedman, J., Carey, V.J., Weiss, S.T., Hohmann, E.L., and Liu, Y.-Y. (2016). Universality of human microbial dynamics. Nature. 534(7606), 259-262.
Bauer, E., Laczny, C.C., Magnusdottir, S., Wilmes, P., and Thiele, I. (2015). Phenotypic differentiation of gastrointestinal microbes is reflected in their encoded metabolic repertoires. Microbiome. 3(1), 1-13. DOI: 10.1186/s40168-015-0121-6.
Bauer, E., Zimmermann, J., Baldini, F., Thiele, I., and Kaleta, C. (2017). BacArena: Individual-based metabolic modeling of heterogeneous microbes in complex communities. PLOS Computational Biology. 13(5), e1005544.
Belenguer, A., Duncan, S.H., Calder, A.G., Holtrop, G., Louis, P., Lobley, G.E., and Flint, H.J. (2006). Two routes of metabolic cross-feeding between Bifidobacterium adolescentis and butyrate-producing anaerobes from the human gut. Applied and environmental microbiology. 72(5), 3593-3599.
Belluzzi, A., Brignola, C., Campieri, M., Pera, A., Boschi, S., and Miglioli, M. (1996). Effect of an enteric-coated fish-oil preparation on relapses in Crohn's disease. New England Journal of Medicine. 334(24), 1557-1560.
Biggs, M.B., Medlock, G.L., Kolling, G.L., and Papin, J.A. (2015). Metabolic network modeling of microbial communities. Wiley Interdisciplinary Reviews: Systems Biology and Medicine. 7(5), 317-334.
Bolger, A.M., Lohse, M., and Usadel, B. (2014). Trimmomatic: a flexible trimmer for Illumina sequence data. Bioinformatics. btu170.
Clausen, M.R., and Mortensen, P. (1995). Kinetic studies on colonocyte metabolism of short chain fatty acids and glucose in ulcerative colitis. Gut. 37(5), 684-689.
De Preter, V., Joossens, M., Ballet, V., Shkedy, Z., Rutgeerts, P., Vermeire, S., and Verbeke, K. (2013). Metabolic profiling of the impact of oligofructose-enriched inulin in Crohn's disease patients: a double-blinded randomized controlled trial. Clinical and translational gastroenterology. 4(1), e30.

den Besten, G., Lange, K., Havinga, R., van Dijk, T.H., Gerding, A., van Eunen, K., Müller, M., Groen, A.K., Hooiveld, G.J., and Bakker, B.M. (2013a). Gut-derived short-chain fatty acids are vividly assimilated into host carbohydrates and lipids. American Journal of Physiology-Gastrointestinal and Liver Physiology. 305(12), G900-G910.

den Besten, G., van Eunen, K., Groen, A.K., Venema, K., Reijngoud, D.-J., and Bakker, B.M. (2013b). The role of short-chain fatty acids in the interplay between diet, gut microbiota, and host energy metabolism. Journal of lipid research. 54(9), 2325-2340.

Dixon, P. (2003). VEGAN, a package of R functions for community ecology. Journal of Vegetation Science. 14(6), 927-930.

Donohoe, D.R., Garge, N., Zhang, X., Sun, W., O'Connell, T.M., Bunger, M.K., and Bultman, S.J. (2011). The microbiome and butyrate regulate energy metabolism and autophagy in the mammalian colon. Cell metabolism. 13(5), 517-526.

Flint, H.J., Bayer, E.A., Rincon, M.T., Lamed, R., and White, B.A. (2008). Polysaccharide utilization by gut bacteria: potential for new insights from genomic analysis. Nature Reviews Microbiology. 6(2), 121-131.

Furusawa, Y., Obata, Y., Fukuda, S., Endo, T.A., Nakato, G., Takahashi, D., Nakanishi, Y., Uetake, C., Kato, K., and Kato, T. (2013). Commensal microbe-derived butyrate induces the differentiation of colonic regulatory T cells. Nature. 504(7480), 446-450.

Gelius-Dietrich, G., Desouki, A.A., Fritzemeier, C.J., and Lercher, M.J. (2013). sybil–Efficient constraint-based modelling in R. BMC systems biology. 7(1), 125.

Griffiths, A.M., Ohlsson, A., Sherman, P.M., and Sutherland, L.R. (1995). Meta-analysis of enteral nutrition as a primary treatment of active Crohn's disease. Gastroenterology. 108(4), 1056-1067.

Guarner, F., and Malagelada, J.-R. (2003). Gut flora in health and disease. The Lancet. 361(9356), 512-519.

Hove, H., and Mortensen, P.B. (1995). Influence of intestinal inflammation (IBD) and small and large bowel length on fecal short-chain fatty acids and lactate. Digestive diseases and sciences. 40(6), 1372-1380.

Huda-Faujan, N., Abdulamir, A., Fatimah, A., Anas, O.M., Shuhaimi, M., Yazid, A., and Loong, Y. (2010). The impact of the level of the intestinal short chain fatty acids in inflammatory bowel disease patients versus healthy subjects. The open biochemistry journal. 4, 53.

Human Microbiome Project Consortium (2012). Structure, function and diversity of the healthy human microbiome. Nature. 486(7402), 207-214. Published online 2012/06/16 DOI: nature11234 [pii]
10.1038/nature11234.

Ji, B., and Nielsen, J. (2015). From next-generation sequencing to systematic modeling of the gut microbiome. Frontiers in genetics. 6.

Kaakoush, N.O., Day, A.S., Huinao, K.D., Leach, S.T., Lemberg, D.A., Dowd, S.E., and Mitchell, H.M. (2012). Microbial dysbiosis in pediatric patients with Crohn's disease. Journal of clinical microbiology. 50(10), 3258-3266.

Kaakoush, N.O., Day, A.S., Leach, S.T., Lemberg, D.A., Nielsen, S., and Mitchell, H.M. (2015). Effect of exclusive enteral nutrition on the microbiota of children with newly diagnosed Crohn's disease. Clinical and translational gastroenterology. 6(1), e71.

Karlsson, F.H., Nookaew, I., and Nielsen, J. (2014). Metagenomic data utilization and analysis (MEDUSA) and construction of a global gut microbial gene catalogue. PLoS computational biology. 10(7), e1003706.

Khor, B., Gardet, A., and Xavier, R.J. (2011). Genetics and pathogenesis of inflammatory bowel disease. Nature. 474(7351), 307-317.

Koropatkin, N.M., Cameron, E.A., and Martens, E.C. (2012). How glycan metabolism shapes the human gut microbiota. Nature Reviews Microbiology. 10(5), 323-335.

Lewis, J.D., Chen, E.Z., Baldassano, R.N., Otley, A.R., Griffiths, A.M., Lee, D., Bittinger, K., Bailey, A., Friedman, E.S., and Hoffmann, C. (2015). Inflammation, antibiotics, and diet as environmental stressors of the gut microbiome in pediatric Crohn's disease. Cell host & microbe. 18(4), 489-500.

Li, H., and Durbin, R. (2009). Fast and accurate short read alignment with Burrows–Wheeler transform. Bioinformatics. 25(14), 1754-1760.

Li, H., Handsaker, B., Wysoker, A., Fennell, T., Ruan, J., Homer, N., Marth, G., Abecasis, G., and Durbin, R. (2009). The sequence alignment/map format and SAMtools. Bioinformatics. 25(16), 2078-2079.

Lien, K.A., McBurney, M.I., Beyde, B.I., Thomson, A., and Sauer, W.C. (1996). Ileal recovery of nutrients and mucin in humans fed total enteral formulas supplemented with soy fiber. The American journal of clinical nutrition. 63(4), 584-595.

Machiels, K., Joossens, M., Sabino, J., De Preter, V., Arijs, I., Eeckhaut, V., Ballet, V., Claes, K., Van Immerseel, F., and Verbeke, K. (2013). A decrease of the butyrate-producing species Roseburia hominis and Faecalibacterium prausnitzii defines dysbiosis in patients with ulcerative colitis. Gut. gutjnl-2013-304833.

Magnusdottir, S., Heinken, A., Kutt, L., Ravcheev, D.A., Bauer, E., Noronha, A., Greenhalgh, K., Jager, C., Baginska, J., Wilmes, P., et al. (2016). Generation of genome-scale metabolic reconstructions for 773 members of the human gut microbiota. Nat Biotech. advance online publication. DOI: 10.1038/nbt.3703

Manichanh, C., Rigottier-Gois, L., Bonnaud, E., Gloux, K., Pelletier, E., Frangeul, L., Nalin, R., Jarrin, C., Chardon, P., and Marteau, P. (2006). Reduced diversity of faecal microbiota in Crohn's disease revealed by a metagenomic approach. Gut. 55(2), 205-211.

Maxwell, E.G., Belshaw, N.J., Waldron, K.W., and Morris, V.J. (2012). Pectin–an emerging new bioactive food polysaccharide. Trends in Food Science & Technology. 24(2), 64-73.

Nookaew, I., Olivares-Hernández, R., Bhumiratana, S., and Nielsen, J. (2011). Genome-scale metabolic models of Saccharomyces cerevisiae. Yeast Systems Biology: Methods and Protocols. 445-463.

O'brien, E.J., Lerman, J.A., Chang, R.L., Hyduke, D.R., and Palsson, B.Ø. (2013). Genome-scale models of metabolism and gene expression extend and refine growth phenotype prediction. Molecular systems biology. 9(1), 693.

Orth, J.D., Thiele, I., and Palsson, B.Ø. (2010). What is flux balance analysis? Nature biotechnology. 28(3), 245-248.

Pamer, E. (2014). Fecal microbiota transplantation: effectiveness, complexities, and lingering concerns. Mucosal immunology. 7(2), 210-214.

Peng, L., He, Z., Chen, W., Holzman, I.R., and Lin, J. (2007). Effects of butyrate on intestinal barrier function in a Caco-2 cell monolayer model of intestinal barrier. Pediatric research. 61(1), 37-41.

Prantera, C., Zannoni, F., Scribano, M.L., Berto, E., Andreoli, A., Kohn, A., and Luzi, C. (1996). An antibiotic regimen for the treatment of active Crohn's disease: a randomized, controlled clinical trial of metronidazole plus ciprofloxacin. American Journal of Gastroenterology. 91(2).

Qin, J., Li, R., Raes, J., Arumugam, M., Burgdorf, K.S., Manichanh, C., Nielsen, T., Pons, N., Levenez, F., and Yamada, T. (2010). A human gut microbial gene catalogue established by metagenomic sequencing. Nature. 464(7285), 59-65.

Roediger, W. (1982). Utilization of nutrients by isolated epithelial cells of the rat colon. Gastroenterology. 83(2), 424-429.

Sabatino, A., Morera, R., Ciccocioppo, R., Cazzola, P., Gotti, S., Tinozzi, F., Tinozzi, S., and Corazza, G. (2005). Oral butyrate for mildly to moderately active Crohn's disease. Alimentary pharmacology & therapeutics. 22(9), 789-794.

Sengupta, S., Muir, J.G., and Gibson, P.R. (2006). Does butyrate protect from colorectal cancer? Journal of gastroenterology and hepatology. 21(1), 209-218.

Thiele, I., Heinken, A., and Fleming, R.M. (2013a). A systems biology approach to studying the role of microbes in human health. Current opinion in biotechnology. 24(1), 4-12.

Thiele, I., and Palsson, B.Ø. (2010). A protocol for generating a high-quality genome-scale metabolic reconstruction. Nature protocols. 5(1), 93-121.
Thiele, I., Swainston, N., Fleming, R.M., Hoppe, A., Sahoo, S., Aurich, M.K., Haraldsdottir, H., Mo, M.L., Rolfsson, O., and Stobbe, M.D. (2013b). A community-driven global reconstruction of human metabolism. Nature biotechnology. 31(5), 419-425.
Tremaroli, V., and Bäckhed, F. (2012). Functional interactions between the gut microbiota and host metabolism. Nature. 489(7415), 242-249.
Trosvik, P., and Muinck, E.J. (2015). Ecology of bacteria in the human gastrointestinal tract—identification of keystone and foundation taxa. Microbiome. 3(1), 44.
Van Dullemen, H.M., van Deventer, S.J., Hommes, D.W., Bijl, H.A., Jansen, J., Tytgat, G.N., and Woody, J. (1995). Treatment of Crohn's disease with anti-tumor necrosis factor chimeric monoclonal antibody (cA2). Gastroenterology. 109(1), 129-135.
Wilcoxon, F. (1945). Individual comparisons by ranking methods. Biometrics bulletin. 1(6), 80-83.
Wilschanski, M., Sherman, P., Pencharz, P., Davis, L., Corey, M., and Griffiths, A. (1996). Supplementary enteral nutrition maintains remission in paediatric Crohn9s disease. Gut. 38(4), 543-548.
Zoetendal, E., Rajilić-Stojanović, M., and De Vos, W. (2008). High-throughput diversity and functionality analysis of the gastrointestinal tract microbiota. Gut. 57(11), 1605-1615.

# Methods

## Key resources table

| REAGENT or RESOURCE | SOURCE | IDENTIFIER |
|---|---|---|
| Biological Samples | | |
| Metagenomic samples of healthy controls and Crohn's disease patients | Lewis et al., 2015 | SRP057027 |
| Deposited Data | | |
| Human genome version 38 | Genome Reference Consortium | http://www.ncbi.nlm.nih.gov/projects/genome/assembly/grc/ |
| 773 metabolic reconstructions (AGORA) | Magnusdottir et al., 2016 | http://vmh.uni.lu/#downloadview |
| Software and Algorithms | | |

| Trimmomatic | Bolger et al., 2014 | http://www.usadellab.org/cms/?page=trimmomatic |
|---|---|---|
| Burrows-Wheeler Aligner (BWA) | Li et al., 2009 | https://sourceforge.net/projects/bio-bwa/files/ |
| Samtools | Li et al., 2009 | https://sourceforge.net/projects/samtools/files/ |
| ILOG CPLEX | IBM | https://www.ibm.com/bs-en/marketplace/ibm-ilog-cplex |
| R | The R Foundation | https://www.r-project.org/ |
| R package BacArena | Bauer et al., 2017 | https://cran.r-project.org/web/packages/BacArena/index.html |
| R package vegan | Oksanen et al., 2009 | https://cran.r-project.org/web/packages/vegan/index.html |
| R package sybil | Gelius-Dietrich et al., 2013 | https://cran.r-project.org/web/packages/sybil/index.html |

## Contact for reagent and resource sharing

Further information and requests for resources and software should be directed to the Lead Contact, Ines Thiele (ines.thiele@uni.lu).

# Experimental model and subject details

Paired-end Illumina raw reads of a study on early onset Crohn's disease (CD) patients and healthy controls of a North American cohort (Lewis et al., 2015) were retrieved from NCBI SRA under the accession: SRP057027. Based on the studies' definition of healthy and dysbiotic individual microbiotas (Lewis et al., 2015), the samples were selected to a smaller subset of 26 healthy controls and 28 CD patients to capture the most pronounced differences in the individual microbial communities. The healthy controls encompassed 12 females and the average age was 14.3 years. The CD patients encompassed 13 females and the average age was 13.9 years. Read accessions and ids of the used samples can be found in Table S2.

# Method details

### Retrieval of metagenomic data and pre-processing

Paired-end Illumina raw reads of a study on early onset Crohn's disease (CD) patients and healthy controls of a North American cohort (Lewis et al., 2015) were retrieved from NCBI SRA under the accession: SRP057027. Based on the studies' definition of healthy and dysbiotic individual microbiotas (Lewis et al., 2015), the samples were selected to a smaller subset of 26 healthy controls and 28 CD patients to capture the most pronounced differences in the individual microbial communities. The selected patients showed pronounced differences in their functional and microbial profile in the original study (Lewis et al., 2015). Furthermore, only the first measured time point was selected to represent newly diagnosed and yet untreated microbiotas. The reads were quality trimmed using Trimmomatic (Bolger et al., 2014) with default parameters for paired-end Illumina sequences. To remove human contaminant sequences, the reads which were still paired after the quality control were mapped with default parameters using the software BWA (Li and Durbin, 2009) to the human genome version 38 (http://www.ncbi.nlm.nih.gov/projects/genome/assembly/grc/).

### Metagenomic mapping and abundance estimation

Using BWA (Li and Durbin, 2009), the pre-processed reads were mapped with default parameters onto a reference set of 773 genomes, which were selected according to a previous study (Magnusdottir et al., 2016) (see also below). Before mapping, the reference genomes of these organisms were combined into one file where each genome is represented as a chromosome. To filter out cross-mapped reads (reads mapped to multiple positions), samtools (Li et al., 2009) was used to discard mapped reads with a low-quality score. The coverage per genome (number of mapped reads normalized by genome size) was calculated using samtools. To reduce the number of false positives, we set a threshold of at least 1% genome coverage for each microbe in each human individual. In accordance to another pipeline (Karlsson et al., 2014), the resulting coverages were normalized for each individual to obtain the relative microbe abundances.

## Microbial metabolic reconstructions

We retrieved published gut microbial metabolic reconstruction (Magnusdottir et al., 2016) from http://vmh.life. These microbes have been chosen according to their prevalence in the human gut and the availability of a genome sequences, and they have been extensively curated based on available physiological and biochemical data (Magnusdottir et al., 2016). In average, a microbial reconstruction consisted of 933 +/- 139 metabolites, 1,198 +/- 241 reactions, and 771 +/- 262 genes.

## Analysis of mapped abundance and reaction differences

The mapped microbial abundances for each individual were compared by computing the Bray-Curtis similarity and subsequent visualization with principal coordinate analysis (PCoA) using the R package vegan (Dixon, 2003). The unique reaction set of personalized *in silico* microbiota was determined by taking the union of all present microbe reactions. These reactions were retrieved from the corresponding metabolic models (Magnusdottir et al., 2016) of each microbe. PCoA was performed on the metabolic distance between each individual's reaction set similar to (Magnusdottir et al., 2016).

## Setup, integration, and simulation of the personalized microbiota models

In the previous steps, the microbial abundance information for each individual was determined. The next step is to integrate this information into a personalized *in silico* microbiota for each person. Therefore, we used a previously established R package for community modeling (Bauer et al., 2017), which represents bacteria as individuals in a grid environment that can exchange metabolites by secretion and uptake. Multiple species can be integrated into the environment with varying number of individuals per species. The dimensions of the two-dimensional quadratic environment was set 0.025 cm$^2$ with 100 grid cells per side length. This resulted in 10,000 grid cells that could be potentially occupied by the microbes. To allow space for the *in silico* microbial community to grow, only 500 microbes were initially added to the grid environment. The relative microbial abundances were used to determine the number of microbes to be added per species (e.g., if one species has a relative abundance 0f 0.01, 5 microbes were added for this species on the entire grid). In case the calculated number of microbes resulted in decimal places, we rounded the final number to the next highest integer. All possible metabolites (union of metabolites that can be taken up by each microbe) were added to the environment with a minimal concentration of 0.2 µM to provide a rich medium that is consistent between individuals. Therefore, metabolite concentrations that emerge from the simulations can be specifically attributed to the microbiota of each individual.

Once the *in silico* microbiota for each CD patient and healthy control have been setup in BacArena, the growth of each microbial model in the microbiota was sequentially for each time step. A total of 24 time steps were simulated, one per hour, corresponding to an overall simulation time of 24 hours. At each time step, the medium composition of each grid cell was updated as a function of the metabolites that were taken up or secreted by the occupying microbes. When a certain growth rate of a microbe occupying a grid cell was reached, a neighboring free grid cell could be occupied by the microbe. If no neighboring grid cell was available, then cells do not duplicate. To reduce the complexity of the model, we simulated a well-mixed environment in which metabolite concentrations are uniformly distributed and microbes move randomly.

The R package sybil (Gelius-Dietrich et al., 2013) was used for constraint based modeling. ILOG CPLEX was used as a linear programming solver. The computations were carried out on high performance computer clusters.

## Analysis of simulation results

After the simulation, each personalized *in silico* microbiota was primarily analyzed in terms of the microbe abundance and metabolite concentrations. Since the simulations include temporal dynamics with different time points, we chose the last time point (24h) for our analysis and comparison between individuals. This allowed the *in silico* microbial communities enough time to consume and produce metabolites, and to reach a steady state. The microbial abundances were determined by assessing the number of microbes in each personalized *in silico* microbiota. The vector of microbial abundances was then compared by computing the Bray-Curtis similarity and subsequently visualized with PCoA. Abundances of specific taxa were calculated by summing up the relative abundances of each corresponding representative. The abundances of the most differing taxa were tested for significant differences between healthy controls and CD patients with the Wilcoxon rank-sum test (Wilcoxon, 1945).

Metabolite concentrations were determined by their molar concentration in the environment at the end of the simulation (t=24h). The concentration of the most relevant metabolites, butyrate, propionate, isobutyrate, L-lactate, and acetate, were assessed and tested for significant differences between the personalized *in silico* microbiota of healthy controls and of CD patients using the Wilcoxon rank-sum test. To investigate the influence of each microbial taxa on the metabolite concentrations, we further evaluated the metabolic fluxes of each microbe in the personalized *in silico* microbiota. For each taxa, the reaction fluxes in all corresponding microbes were summed up.

## Definition of personalized dietary treatments

After identifying the metabolic signatures influencing CD and the corresponding microbes causing these differences between healthy controls and CD patients, we predicted metabolites that could revert these differences. Therefore, we analyzed each genome-scale microbial metabolic model separately.

According to their presence in each personalized *in silico* microbiota, the set of microbes was selectively analyzed for every individual. Each personalized *in silico* microbiota was then simulated in a rich medium containing all possible metabolite with flux uptake constraints of 1 mmol gDW$^{-1}$ h$^{-1}$ and the biomass as well as the production of SCFAs (butyrate, propionate, isobutyrate, L-lactate, acetate) were optimized for separately. To enhance the growth of beneficial bacteria, we selected metabolites based on the ability of the CD low abundant microbes (e.g., Clostridia, Bacteroides) to uptake these nutrients over the CD high abundant microbes (e.g., Gammaproteobacteria, Bacilli). We then added the selected metabolites iteratively to the *in silico* medium with a maximal flux uptake constraint of 1000 mmol gDW$^{-1}$ h$^{-1}$ to investigate whether the fermentation products increased or decreased. Based on these simulations, the added metabolites which had a positive effect (recovering metabolite production to healthy levels) were then collected and used as the personalized dietary treatment for each individual.

We tested the effect of the treatment on the personalized *in silico* microbiota of CD patients by adding a 100 times higher concentration of the predicted treatment metabolites to the *in silico* rich diet containing 0.2 µM for each metabolite. The personalized *in silico* microbiota simulations and analyses were then carried out as described above.

## Quantification and statistical analysis

Differences between healthy controls and CD patients were assessed with the Wilcoxon rank-sum test implemented in R. For the healthy controls our group size was 26 individuals and for the CD group 28 individuals.

## Data and software availability

The scripts to construct and simulate the individual specific microbiota models as well as the analysis scripts are available on GitHub: https://github.com/euba/CodeBase